\documentstyle[wstwocl,epsfig]{article}
\begin{document}

\bibliographystyle{unsrt}    

\newcommand{\st}{\scriptstyle}
\newcommand{\sst}{\scriptscriptstyle}
\newcommand{\mco}{\multicolumn}
\newcommand{\epp}{\epsilon^{\prime}}
\newcommand{\vep}{\varepsilon}
\newcommand{\ra}{\rightarrow}
\newcommand{\ppg}{\pi^+\pi^-\gamma}
\newcommand{\vp}{{\bf p}}
\newcommand{\ko}{K^0}
\newcommand{\kb}{\bar{K^0}}
\newcommand{\al}{\alpha}
\newcommand{\ab}{\bar{\alpha}}
\def\be{\begin{equation}}
\def\ee{\end{equation}}
\def\bea{\begin{eqnarray}}
\def\eea{\end{eqnarray}}
\def\CPbar{\hbox{{\rm CP}\hskip-1.80em{/}}}

\def\ap#1#2#3   {{\em Ann. Phys. (NY)} {\bf#1} (#2) #3.}
\def\apj#1#2#3  {{\em Astrophys. J.} {\bf#1} (#2) #3.}
\def\apjl#1#2#3 {{\em Astrophys. J. Lett.} {\bf#1} (#2) #3.}
\def\app#1#2#3  {{\em Acta. Phys. Pol.} {\bf#1} (#2) #3.}
\def\ar#1#2#3   {{\em Ann. Rev. Nucl. Part. Sci.} {\bf#1} (#2) #3.}
\def\cpc#1#2#3  {{\em Computer Phys. Comm.} {\bf#1} (#2) #3.}
\def\err#1#2#3  {{\it Erratum} {\bf#1} (#2) #3.}
\def\ib#1#2#3   {{\it ibid.} {\bf#1} (#2) #3.}
\def\jmp#1#2#3  {{\em J. Math. Phys.} {\bf#1} (#2) #3.}
\def\ijmp#1#2#3 {{\em Int. J. Mod. Phys.} {\bf#1} (#2) #3.}
\def\jetp#1#2#3 {{\em JETP Lett.} {\bf#1} (#2) #3.}
\def\jpg#1#2#3  {{\em J. Phys. G.} {\bf#1} (#2) #3.}
\def\mpl#1#2#3  {{\em Mod. Phys. Lett.} {\bf#1} (#2) #3.}
\def\nat#1#2#3  {{\em Nature (London)} {\bf#1} (#2) #3.}
\def\nc#1#2#3   {{\em Nuovo Cim.} {\bf#1} (#2) #3.}
\def\nim#1#2#3  {{\em Nucl. Instr. Meth.} {\bf#1} (#2) #3.}
\def\np#1#2#3   {{\em Nucl. Phys.} {\bf#1} (#2) #3.}
\def\pcps#1#2#3 {{\em Proc. Cam. Phil. Soc.} {\bf#1} (#2) #3.}
\def\pl#1#2#3   {{\em Phys. Lett.} {\bf#1} (#2) #3.}
\def\prep#1#2#3 {{\em Phys. Rep.} {\bf#1} (#2) #3.}
\def\prev#1#2#3 {{\em Phys. Rev.} {\bf#1} (#2) #3.}
\def\prl#1#2#3  {{\em Phys. Rev. Lett.} {\bf#1} (#2) #3.}
\def\prs#1#2#3  {{\em Proc. Roy. Soc.} {\bf#1} (#2) #3.}
\def\ptp#1#2#3  {{\em Prog. Th. Phys.} {\bf#1} (#2) #3.}
\def\ps#1#2#3   {{\em Physica Scripta} {\bf#1} (#2) #3.}
\def\rmp#1#2#3  {{\em Rev. Mod. Phys.} {\bf#1} (#2) #3.}
\def\rpp#1#2#3  {{\em Rep. Prog. Phys.} {\bf#1} (#2) #3.}
\def\sjnp#1#2#3 {{\em Sov. J. Nucl. Phys.} {\bf#1} (#2) #3.}
\def\spj#1#2#3  {{\em Sov. Phys. JEPT} {\bf#1} (#2) #3.}
\def\spu#1#2#3  {{\em Sov. Phys.-Usp.} {\bf#1} (#2) #3.}
\def\zp#1#2#3   {{\em Zeit. Phys.} {\bf#1} (#2) #3.}

\setcounter{secnumdepth}{2} 

   
\title{REVIEW OF HADRONIC AND RARE B DECAYS}

\firstauthors{T.E. Browder}

\firstaddress{Department of Physics and Astronomy, University of
Hawaii, 
Honolulu, HI 96822, U.S.A \\    
University of Hawaii preprint UH 511-836-95 \\
(To appear in the Proceedings of the 1995 Brussels Europhysics
Conference.) }

\twocolumn[\maketitle]

\section{Introduction}

We discuss new results on rare and hadronic B decays from
the CLEO, CDF and LEP experiments.

\section{Inclusive $B$ Decays and Charm Counting}

A complete set of measurements of inclusive $B$ decays is now
emerging from the CLEO experiment. When combined with observations
of lepton-charmed particle correlations these will provide a full picture
of the weak interaction mechanisms which operate in $B$ decay.
The measurements of branching fractions for inclusive $B$ decays
are given in Table~\ref{khinc}.

Preliminary results on $B\to D^0 X$ and $B\to D^+ X$ branching fractions
were shown at this conference.
From these and earlier
results\cite{CLEODsX},\cite{CLEOLc},\cite{CLEOSc},
\cite{CLEOcc},\cite{BH}, we can obtain $n_c$, the number of
charm quarks produced per $B$ decay. I have calculated the value of
$n_c$ using measurements from CLEO~II alone and using
world averages\cite{ncdetails}. 
Since many of the inclusive measurements
are limited by the systematic uncertainties in charmed meson
absolute branching fractions, the CLEO~II results do not completely
dominate the world average.
In Table~\ref{khtotal}, we give the value of $n_c$ for the CLEO~II
experiment. We also a compute a world average for
$n_c$ by combining measurements
from the CLEO~1.5, ARGUS and CLEO~II experiments.

\begin{table}[h] 
\caption{Total Charm Yield in $B$ Decay [\%]}
\label{khtotal} 
 \begin{center} 
\begin{tabular}{ll}
\hline\hline 
 CLEO II & World Average \\ 
\hline 
  $ 115.8 \pm 5.26 $ & $ 111.7 \pm 4.56 $ \\ 
\hline\hline 
\end{tabular} 
\end{center} 
\end{table} 

\begin{table}[h] 
\caption{$b \to c\bar{c}s$ Fraction [\%]}
\label{khccs} 
 \begin{center} 
\begin{tabular}{ll}
\hline\hline  
 CLEO II & World Average \\ 
\hline 
  $ 17.9 \pm 2.02 $ & $ 16.1 \pm 1.76 $ \\ 
\hline\hline 
\end{tabular} 
\end{center} 
\end{table}

Particle-lepton correlations can give important
information on the production
mechanisms operative in inclusive $B$ decay. For instance
last year, motivated by the low charm yield in inclusive $B$ decay,
 the possibility of a new mechanism operative
in $B\to$~baryon decay was suggested by Falk, Wise, and Dunietz\cite{Falk}.
The mechanism,
internal $W$ emission in $b\to c \bar{c} s$ decay, gives
rise to $\Lambda_c^+ - \ell^+$ correlations whereas the
usual $b\to c \bar{u} d$ mechanisms give 
the opposite $\Lambda_c^+ - \ell^-$
correlations. Note that the charge of the high momentum lepton is used
to tag the flavor of the other B meson.
A modest signal is observed in the wrong sign
$\Lambda_c^+ - \ell^+$ correlation indicating that
the ratio $b\to c \bar{c} s/b\to c \bar{u} d$ in B$\to$ baryon
decays is $0.20\pm 0.13\pm 0.04$\cite{CLEOSc}.
Similiarly the possibility of $B\to D_s^+$ production
from $b\to c \bar{u} d$ processes with $s \bar{s}$ quark popping
can be distinguished from direct $B\to D_s^+$  production from
the quark level process $b\to c \bar{c} s$ by the sign of
the $D_s-\ell$ correlation. 
A limit on the former possibility 
(i.e. $b\to c \bar{u} d$ with $s \bar{s}$ quark popping)
 can be deduced from the upper limit  reported
on $D_s^+ -\ell^+$ correlations at this conference
($<31\%$ of the total)\cite{dslep}.

\setlength{\unitlength}{0.4mm}
\begin{figure}[hbt]
\begin{picture}(180,190)(-35,305)
\mbox{\epsfxsize16.0cm\epsffile{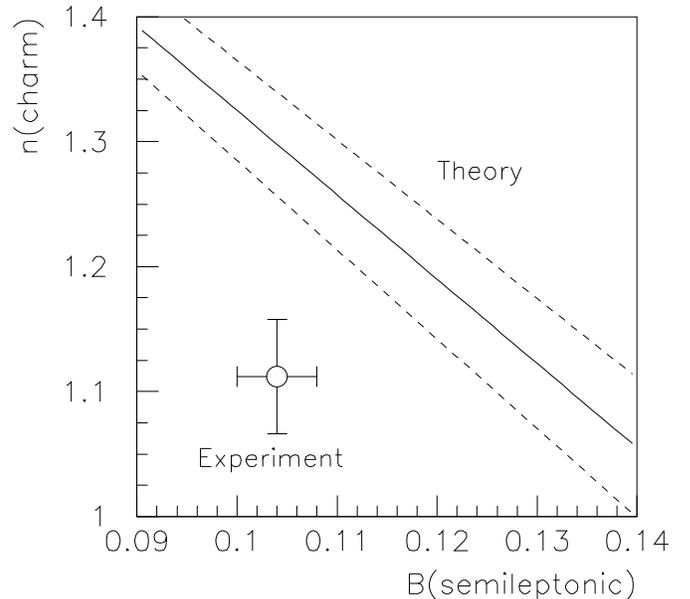}}
\end{picture}
\caption{The charm yield per $B$ decay versus the
B semileptonic branching fraction. The point with error
bars is the experimental world average. The solid line
is the theoretical prediction. The dashed lines are
upper bounds on the theoretical uncertainties.}
\end{figure}

In Table~\ref{khccs}, we also give ${\cal B}(b\to c\bar{c} s)$
which is computed 
from the observed rates for $B\to D_s$, $B\to \psi$, $B\to \psi^{'}$,
and $B\to \Xi_c$ decays. The branching fraction is below $30\%$, 
the value which
is theoretically required to accomodate
the low value of the semileptonic branching 
fraction\cite{Russkie},\cite{Falk},\cite{ISI}.
The possible contribution of $B\to D\bar{D} K$
decays, which corresponds to the quark level process $b\to c \bar{c} s$
with popping of a light quark pair, is not included. 
Buchalla, Dunietz, and Yamamoto have
recently suggested that the latter mechanism
may be significant\cite{ISI}. 
This possibility leads to wrong sign
$D-\ell$ correlations and
is currently under investigation at CLEO\cite{Kwon}.

Figure 1 shows a comparison of the world average value of
$n_c$ and the theoretical expectation from the parton model\cite{BBSL}.
For the world average of the $B$ semileptonic branching fraction,
the expectation is $n_c\approx 1.3$.
Thus, theory and experiment disagree. The origin of this discrepancy
is still not well understood. A number of explanations have been
suggested.

One possibility is a
systematic flaws in the experimental determination of the yield of
charm quarks.
The charm meson absolute branching
fractions can contribute a systematic uncertainty, although the 
errors from this source
 have been significantly reduced by the precise recent determinations
of ${\cal B}(D^0\to K^-\pi^+)$
and ${\cal B}(D^+\to K^-\pi^+\pi^+)$. However, the absolute branching
fraction scales for the $D_s$ meson and $\Lambda_c$ baryons are
still quite uncertain. 
Since the inclusive branching ratios to
these particles are small, a substantial change to the branching ratio
scale would be required to significantly modify the charm yield.

Another possibility is
a large production mechanism that contributes 
to the inclusive rate but that has not been measured or identified.
Palmer and Stech have suggested\cite{palmstech},
that $b \to c \bar{c} s$ followed by $c \bar{c} \to \rm{gluons}$,
 which in turn hadronize into a final state with no charm, has a large
branching ratio. The charm content for this mechanism would not be properly
taken into account.
Another related suggestion is that the rate for the hadronic penguin
diagram $b\to s g$ is larger than expected\cite{kaganbsg}.

\section{Color suppressed amplitude}
We now discuss progress on determination of the strength and
magnitude of the internal W emission or
``color suppressed'' amplitude. The evidence for the presence
of this mechanism are $B\to \psi$ decays which result from internal
W-emission ($b\to c \bar{c} s$). To date, there is no direct observation
of the corresponding internal W-emission process $b\to c\bar{u} d$ which
results in final states such as $\bar{B}^0\to D^{(*)0} \pi^0$ and
$\bar{B}^0\to D^{(*)0} \eta^{(')}$.
Improved upper limits on these
decay modes were reported at this conference\cite{CLEOcolor}. The best limit  
$${\cal B}(\bar{B}^0\to D^0\pi^0)/{\cal B}(\bar{B}^0\to D^+\pi^-)<0.11$$
is near the range expected from models ($0.03-0.08$).
 Measurements of $B^-$ decay modes,
where both external and internal W-emission amplitudes contribute, give
indirect information on the size of the color suppressed amplitude in
$b\to c\bar{u} d$ and show that it has a positive sign\cite{BH}.

Improved measurements
of decays involving charmonia were reported at this conference
by CLEO and CDF. A measurement of the
polarization in their large and clean sample of
reconstructed $\bar{B}^0\to \psi K^{*0}$ events
has been completed by CDF. They find
$${\Gamma_L}/{\Gamma_T}= 0.65\pm 0.10\pm 0.04~({\rm CDF}).$$
This is consistent with but slightly lower than the two previous measurements
from ARGUS and CLEO~II. A new world average can also be calculated from 
these measurements,  
$${\Gamma_L}/{\Gamma_T}= 0.78\pm 0.07 ~(\rm{World~Average}).$$
In addition, CDF reported the first measurement of polarization
for the $B_s\to \psi\phi$ mode,
$${\Gamma_L}/{\Gamma_T}= 0.56\pm{0.21}^{+0.02}_{-0.04}~({\rm CDF}).$$

The CDF experiment also reported observation of a Cabibbo suppressed
decay mode $B^+\to \psi \pi^+$. They found $25\pm 8$ events and performed
a number of consistency checks to verify that these events did not originate
from the kinematic reflection of $B^+\to \psi K^+$. 
The ratio of branching fractions\cite{cdfpsipi}
$${\cal B}(B^+\to \psi\pi^+)/{\cal B}(B^+\to \psi K^+)=
4.9^{+1.9}_{-1.7}\pm 1.1\%$$
is consistent with an earlier CLEO observation and the updated result
$${\cal B}(B^+\to \psi\pi^+)/{\cal B}(B^+\to \psi K^+)=
5.2\pm 2.4\%$$
presented at this conference\cite{cleopsipi}.

The factorization hypothesis works at the present 
level of experimental
precision for $B\to D^{*}\pi^-$, $B\to D^{*}\rho^-, D^{*} a_1^-$ decays
which are 
external W-emission $b\to c \bar{u} d$ processes with large energy 
release.
It is difficult, however to simultaneously accomodate the value\cite{BH}
$$R={\cal B}(B\to \psi K^*)/{\cal B}(B\to \psi K)=1.68\pm 0.33$$
and the observed polarization in $B\to \psi K^*$ in models based
on factorization\cite{aleksan},\cite{kamal}. 
It is open question as to whether this indicates
that factorization breaks down in internal W-emission or 
that there is some other flaw
in the models.

\section{Rare $B$ Decays and Gluonic Penguins}
The CLEO experiment has recently presented a measurement
of the inclusive $b\to s \gamma$ branching fraction. So far, 
no unambiguous evidence for the gluonic analogue,
$b\to s$~gluon, has been
found. 

CLEO has recently updated their measurement of
the sum of the branching fractions for ${B^0}\to \pi^+\pi^-$
and ${B^0}\to K^+ \pi^-$, they find
a signal which contains $17.2^{+5.6+2.2}_{-4.9-2.5}$ events
which corresponds to a branching fraction
$${\cal B}({B}^0\to \pi^+\pi^-, K^+\pi^-)= (1.8^
{+0.6+0.2}_{-0.5-0.3}\pm 0.2)\times 10^{-5}$$
There is insufficient statistics to claim a signal in either
of the individual modes and upper limits of 
$${\cal B}({B}^0\to \pi^+ \pi^-)< 2.0 \times 10^{-5}$$
$${\cal B}({B}^0\to K^+ \pi^-)< 1.7 \times 10^{-5}$$ are given.
Therefore it is not possible to isolate the contributions of
the spectator and penguin amplitudes.

The LEP experiments have begun to achieve an interesting sensitivity
to rare $B$ decays. Limits on these two body charged modes in the range
$(4-8)\times 10^{-5}$ have been obtained by ALEPH, and OPAL.
The utility of the silicon vertex detectors is demonstrated by
the low backgrounds and sensitivities achieved by ALEPH in 
their search for higher multiplicity rare $B$ decays such as
$\bar{B}\to 2\pi^+ 2\pi^-)$.
DELPHI presented evidence of a signal in the sum of two
body modes. The observed signal, which contains
 3 events with an estimated background of 0.18 events, corresponds to the
branching fraction
$${\cal B}({B}^0\to \pi^+\pi^-, K^+\pi^-)= (2.5\pm 1.4) \times 10^{-5}$$ 
and is consistent with the CLEO measurement.
The L3 experiment take advantage of their electromagnetic calorimetry
to obtain limits on rare $B$ decays with multineutrals (e.g.
$\bar{B}\to \eta\pi^0$). Their limits on modes in this class are in the range 
$(8\times 10^{-5})-(3\times 10^{-4})$.

It is also possible to constrain the penguin amplitude by searching
for $b\to s \bar{s} s$ decays 
such as $B\to \phi K^0$ which do not have contributions
from other processes. CLEO has searched for the decay modes
$\bar{B}\to\phi K^{(*)}$ and obtained limits on their branching fractions
in the range $(1.2-8.8)\times 10^{-5}$.

Recently another method for constraining $b\to s$~gluon using inclusive
processes has been introduced by CLEO. It is possible to examine the
endpoint region in inclusive $\phi$ production beyond the kinematic
limit for $b\to c$ processes. This corresponds roughly to the range 
$2.7> p(\phi) >2.0$ GeV. 

Two experimental methods have been applied\cite{CLEOphi}. One can search for 
an excess in the endpoint region after applying loose event 
shape cuts. This method has a minimal dependence on the model
of the $X_s$ hadronization. No signal is observed.
This method gives an upper limit of
${\cal B}(B\to X_s\phi)< 2.2 \times 10^{-4}$ for $2.7>p(\phi)>2.0$ GeV
An alternate approach is to combine $\phi$ mesons with
n charged pions and a neutral pion, where $n=1-4$, and then 
form the beam constrained mass or energy difference. No attempt
is made to correct for misidentification of decay modes. Then combinations
which are inconsistent with the $B$ mass and the beam energy are discarded
in order to suppress continuum.
This gives an upper limit of ${\cal B}(B\to X_s\phi)< 1.1 \times 10^{-4}$
for $M(X_s)<2.0$ GeV.

To convert these results into limits on 
the total rate ${\cal B}(B\to X_s \phi)$ a theoretical model is required.
We have used the model of Deshpande et al., which implies that most of
the rate is concentrated in quasi two body modes which produce $\phi$ mesons
in the high momentum window\cite{deshbsg}. 
Only a small correction
is required, and the most stringent limit is obtained by the second method
which gives ${\cal B}(B\to X_s\phi)< 1.3 \times 10^{-4}$ at the 
90\% confidence level. More theoretical work is required to understand
the implications for ${\cal B}(b\to s$~gluon).

The inclusive
method of studying rare decays
described here can be extended to $B\to K_s$, 
$B\to K^{*}$, $B\to \eta^{(')}$
transitions in order to search for other signatures
of gluonic penguins.

\section{Conclusion}

A complete picture
of inclusive $B$ decays is emerging from 
recent measurements by the CLEO experiment.
The experimental result for the charm yield per $B$ decay ($n_c$)
is consistent with the naive expectation
that $1.15$ charm quarks are produced per $b$ decay. However,
it does not support the proposals
which suggest that at least $1.3$ quarks should
be produced per $b$ decay. 
Recent theoretical efforts require such a high 
charm yield in order to explain
the discrepancy between 
theoretical calculations and experimental measurements of
the inclusive semileptonic rate, ${\cal B}(B \to X\ell \nu)$ 

The CDF results on B decay modes with charmonia presented at this conference 
provide compelling evidence for the utility of silicon vertex detectors
in $B$ physics and indicate that CDF has now begun to make major contributions
to the study of $B$ {\it decay}.
The theoretical problems of 
interpreting the data and understanding
whether factorization is operative in internal W-emission however remain.

The long struggle to isolate and measure the gluonic penguin amplitude in
$B$ decay continues. This is an essential component of the program of
extracting physics from measurements of CP asymmetries at future facilities.

\setcounter{secnumdepth}{0} 

\section{Acknowledgments}
I thank my colleagues on the CLEO experiment
for their contribution to the work described here. I also thank
the CDF, ALEPH, DELPHI, L3, and OPAL representatives for communicating
their preliminary results. 

\twocolumn{
\begin{table}[hbt] 
\let\tabbodyfont\scriptsize 
\twocolumn[\caption{Branching fractions (\%) of inclusive $B$ decays}]
\medskip
\label{khinc} 
\begin{tabular}{||l|lll|l||} 
\hline\hline
\multicolumn{1}{l}{Particle} & 
\multicolumn{1}{l}{ARGUS} & 
\multicolumn{1}{l}{CLEO 1.5} & 
\multicolumn{1}{l}{CLEO} & 
\multicolumn{1}{l}{Average} \\  
\hline 
 $\bar{B} \rightarrow \bar{D}^0 X$ & $ 49.7 \pm 3.8  \pm 6.4  \pm 2.6  $ & 
$ 59.7 \pm 3.2  \pm 3.6  \pm 3.1  $ & 
$ 64.5 \pm 2.1  \pm 1.4  \pm 1.8  $ & 
$ 61.9 \pm 2.6  $ \\ 
 $\bar{B} \rightarrow D^- X$ & $ 23.0 \pm 3.0  \pm 4.4  \pm 1.5  $ & 
$ 24.9 \pm 3.3  \pm 2.0  \pm 1.6  $ & 
$ 23.5 \pm 1.2  \pm 0.8  \pm 2.3  $ & 
$ 23.8 \pm 2.1  $ \\ 
 $\bar{B} \rightarrow D_s^- X$ & $ 8.3 \pm 1.1  \pm 0.9  \pm 1.0  $ & 
$ 8.7 \pm 1.3  \pm 1.0  $ & 
$ 12.1 \pm 0.4  \pm 0.9  \pm 1.4  $ & 
$ 10.3 \pm 0.7  \pm 1.2  $ \\ 
 $\bar{B} \rightarrow \psi X$ & $ 1.25 \pm 0.19  \pm 0.26  $ & 
$ 1.31 \pm 0.12  \pm 0.27  $ & 
$ 1.12 \pm 0.04  \pm 0.06  $ & 
$ 1.14 \pm 0.07  $ \\ 
 $\bar{B} \rightarrow \psi X$ (direct) & $ 0.95 \pm 0.27  $ & 
 & 
$ 0.81 \pm 0.08  $ & 
$ 0.82 \pm 0.08  $ \\ 
 $\bar{B} \rightarrow \psi$'$ X$ & $ 0.50 \pm 0.19  \pm 0.12  $ & 
$ 0.36 \pm 0.09  \pm 0.13  $ & 
$ 0.34 \pm 0.04  \pm 0.03  $ & 
$ 0.35 \pm 0.05  $ \\ 
 $\bar{B} \rightarrow \chi_{c1} X$ & $ 1.23 \pm 0.41  \pm 0.29  $ & 
 & 
$ 0.40 \pm 0.06  \pm 0.04  $ & 
$ 0.42 \pm 0.07  $ \\ 
 $\bar{B} \rightarrow \chi_{c1} X$ (direct) &  & 
 & 
$ 0.37 \pm 0.07  $ & 
$ 0.37 \pm 0.07  $ \\ 
 $\bar{B} \rightarrow \chi_{c2} X$ &  & 
 & 
$ 0.25 \pm 0.10  \pm 0.03  $ & 
$ 0.25 \pm 0.10  $ \\ 
 $\bar{B} \rightarrow \eta_{c} X$ &  & 
 & 
$ <0.90 $  (90\% C.L.) & 
 $ <0.90 $  (90\% C.L.) \\ 
 $\bar{B} \rightarrow \Lambda _c^- X$ & $ 7.0 \pm 2.8  \pm 1.4  \pm 2.1  $ & 
$ 6.3 \pm 1.2  \pm 0.9  \pm 1.9  $ & 
 & 
$ 6.4 \pm 1.3  \pm 1.9  $ \\ 
  $\bar{B} \rightarrow \Xi_c^+ X$ &  & 
 & 
$ 1.5 \pm 0.7  $ & 
$ 1.5 \pm 0.7  $ \\ 
 $   $ &  &  &  $  $ &  $   $ \\ 
  $\bar{B} \rightarrow \Xi_c^0 X$ &  & 
 & 
$ 2.4 \pm 1.3  $ & 
$ 2.4 \pm 1.3  $ \\ 
 $   $ &  &  &  $  $ &  $   $ \\ 
\hline\hline
\end{tabular} 
\end{table} }



\begin{thebibliography}{99}
\bibitem{Moneti}
The value of the $B\to D^+$ branching fraction has been revised
since the presentation at the Brussels conference. 
Private communication from G. Moneti (CLEO II Collaboration).
\bibitem{ncdetails} For the $B\to D_s X$ measurement, 
the PDG value for the $D_s \to \phi \pi$ absolute
branching fraction (3.5 $\pm$ 0.4) was used. 
To compute $n_c$, the $B\to \Xi_c^o$ result was 
corrected for $s\bar{s}$ popping.
Twice the $\eta_c$ upper limit was used to give a generous 
estimate of the contributions from the unobserved charmonium states.
\end{thebibliography}
\end{document}